\documentclass[twocolumn,prl,showpacs]{revtex4}

\usepackage{graphicx}
\usepackage{amsmath}
\usepackage{amsfonts}
\usepackage{amssymb}
\usepackage{enumerate}
\usepackage{hyperref}
\hypersetup{%
  breaklinks = {true}, 
  colorlinks = {true}, 
  citecolor = {blue},  
  filecolor = {black}, 
  linkcolor = {red},   
  urlcolor  = {blue},  
  pdfauthor = {\textcopyright\ Carsten Raas, Patrick Grete, and G\"otz S. Uhrig},
  pdfstartpage = {1},
  pdfstartview = {FitH}, 
}

\begin{document}

\title{Emergent Collective Modes and Kinks in Electronic Dispersions}

\author{Carsten Raas}
\email{carsten.raas@tu-dortmund.de}
\homepage{http://www.raas.de}
\affiliation{Lehrstuhl f\"{u}r Theoretische Physik I,
  Technische Universit\"{a}t Dortmund, Otto-Hahn Stra\ss{}e 4, 44221 Dortmund, Germany}

\author{Patrick Grete}
\affiliation{Lehrstuhl f\"{u}r Theoretische Physik I,
  Technische  Universit\"{a}t Dortmund, Otto-Hahn Stra\ss{}e 4, 44221 Dortmund, Germany}

\author{G\"otz S. Uhrig}
\email{goetz.uhrig@tu-dortmund.de}
\homepage{http://t1.physik.tu-dortmund.de/uhrig/}
\affiliation{School of Physics,
  University of New South Wales, Kensington 2052, Sydney NSW,
  Australia}
 \altaffiliation{On leave from Lehrstuhl f\"{u}r Theoretische Physik I,
    Technische  Universit\"{a}t Dortmund, Otto-Hahn Stra\ss{}e 4,
    44221 Dortmund, Germany}

\date{\rm\today}

\begin{abstract}
Recently, it was shown that strongly correlated metallic
fermionic systems [Nature Phys. {\bf 3}, 168 (2007)] generically display kinks 
in the dispersion of single fermions without the coupling to collective modes. 
Here we provide compelling evidence that the physical origin of these kinks
are emerging \emph{internal} collective modes of the fermionic systems.
In the Hubbard model under study these modes are identified to be
spin fluctuations which are the precursors of the spin excitations
in the insulating phase. In spite of their damping the emergent modes give rise
to signatures very similar to features of models including
coupling to external modes.
\end{abstract}


  

\maketitle

The description of nascent collective modes which emerge from
elementary excitations on varying a control parameter $g$ is an
intensely studied field of research. The difficulty relies
in the fact that in one limit of $g$ the elementary excitations
dominate while in the other limit the collective modes dominate.
In the vicinity of the transition or around the crossover
necessarily both degrees of freedom need to be taken into account
so that the interplay of both kinds of excitations is crucial.
No simple theory assesses this interplay.

Here we will focus on strongly correlated electronic systems and
especially on the metal-insulator transition induced by
a repulsive interaction $U$ on a lattice with a commensurate number
of electrons per site. The simplest case is a local interaction with
one electron per site on average \cite{georg96}.
For low values of $U$ the electrons
move through the lattice so that the system is metallic. For large
values of $U$ the hopping is blocked and the system is insulating
with frozen charge degree of freedom. But the spin dynamics is still active.
In leading order in $t/U$ ($t$ the hopping matrix element) this
dynamics is captured by a Heisenberg model \cite{harri67}. The collective
modes are the spin excitations built from bound electron-hole pairs.

When the system is still metallic, but close to its insulating regime,
we intend to understand how the emergent spin modes influence the 
electronic quasiparticles. This issue is important to many strongly 
correlated systems. One prominent example is high temperature 
superconductiviy where a large number of theories explains the attractive 
interaction between charge carriers by the interplay with spin fluctuations.
One line of argument links the kinks that are observed in the dispersion
of the fermionic holes, see for instance 
\cite{lanza01,boris06,kordy06,inoso07a,valla07}, 
to the interaction with bosonic modes. This is the usual reasoning for phonons 
coupled to  electrons \cite{scala69}. Other bosonic modes, however,
will engender the same sort of kinks, for instance plasmons \cite{bostw07}.
In the high-$T_c$ materials, spin fluctuations have an important
influence on the quasiparticles, see e.g.\ Ref.\ \cite{guo06}. 
They are likely candidates for the bosonic modes,
 see e.g.\ Ref.\ \cite{mansk01} 
where this is worked out in the fluctuation-exchange approximation.

Byczuk et al.\  \cite{byczu07a} 
recently showed by a sophisticated analysis of the equations of
dynamic mean-field theory (DMFT) \cite{georg96} 
that kinks in the electronic dispersion are a generic feature
of strongly correlated electronic systems where the repulsive interaction
is of similar strength as the kinetic energy. They stress that
no coupling to a bosonic mode is needed. Indeed the model they study
does not comprise any explicit bosonic mode; it is a fermionic
Hubbard model. For particle-hole symmetric models
dominated by the local self-energy the position of the kink
was related by Byczuk et al.\  \cite{byczu07a}  to the quasi-particle 
weight
\begin{equation}
\label{eq:byczuk}
\omega_\text{kink}=(\sqrt{2}-1) Z D.
\end{equation}

In the present work it is our aim to elucidate the physical origin
of the kinks. We provide evidence that the kinks result from 
the coupling to the bosonic resonance which is the precursor of the
spin modes in the insulator. Thus we conjecture that the kinks in 
strongly correlated fermionic systems are induced by coupling to 
\emph{internal} bosonic modes. The signature is very similar to 
the coupling to external bosons such as phonons \cite{scala69}. 
Our finding also sheds 
light on signatures of spin modes in the electronic dispersions
of high-temperature superconductors.

Our computation is also based on DMFT.  This approach reduces the 
extensive lattice problem to a self-consistency problem 
involving a single-impurity Anderson model (SIAM) \cite{georg96}. The latter 
can be viewed as an interacting site coupled to a semi-infinite
chain of non-interacting fermions \cite{hewso93,uhrig96a}
which is solved by dynamic density-matrix renormalization
(D-DMRG) \cite{raas04a,weich08}. This combination of D-DMRG and DMFT represents
a powerful tool for investigating the $T=0$ one-particle propagators of
interacting lattice models [17-20]
Its particular merit is to have a well-controlled energy resolution 
over the whole energy range \cite{gebha03,raas04a}.

The model under study is the simplest displaying
an interaction driven metal-insulator transition, namely the
half-filled Hubbard model
\begin{equation}
\label{eq:hamilton}
\mathcal{H} = -t \sum_{\langle i, j\rangle; \sigma}
 c^\dagger_{i;\sigma}
c^{\phantom\dagger}_{j;\sigma} + U \sum_i (\hat n_{i;\uparrow}-1/2)
(\hat n_{i;\downarrow}-1/2).
\end{equation}
At low values of $U$ the ground state is metallic; above $U_{c2}\approx 3D$
the insulating phase becomes the ground state 
\cite{georg96,bulla99a,karsk05,blume05b}.

Our analysis is facilitated by the direct numerical calculation of the
local proper self-energy $\Sigma(\omega)$. This is done with
the help of the improper self-energy 
\begin{equation}
\label{eq:Q}
Q(\omega):=\langle\langle d_\sigma(\hat n_{-\sigma}-1/2)| 
(\hat n_{-\sigma}-1/2) d_\sigma^\dagger \rangle\rangle
\end{equation}
where we use the notation $\langle\langle A|B\rangle\rangle$
for the Fourier transform of the time-dependent 
fermionic Green function $-i\langle \{ A(t),B(0)\}\rangle$.
If one considers doping the term $1/2$ in \eqref{eq:Q}
is to be replaced by the average filling per site.

Starting from the result $\Sigma(\omega)=U F(\omega)/G(\omega)$ 
by Bulla et al. \cite{bulla98} we apply the Liouville operator 
in the equations of motion once more \cite{fassb05} yielding $F(\omega)=U Q
(\omega) G_0(\omega)$ wherein $F(\omega):=\langle\langle d_\sigma
(\hat n_{-\sigma}-1/2)|  d_\sigma^\dagger \rangle\rangle$. Substituting
$F(\omega)$ by $U Q(\omega) G_0(\omega)$ and expressing $G(\omega)$
by Dyson's equation $G^{-1}(\omega)=G^{-1}(\omega)-\Sigma(\omega)$
yields
\begin{equation}
\label{eq:Sigma}
\Sigma(\omega) = U^2 Q(\omega)/(1+U^2Q(\omega) G_0(\omega)).
\end{equation}
This expression is advantageous to use for small to moderate
values of $U\lessapprox 2D$ where the computation of $\Sigma(\omega)$ 
from the difference between the inverse bare and full propagators
is numerically not reliable \cite{karsk08}.

\begin{figure}
\begin{center}
     \includegraphics[width=0.82\columnwidth]{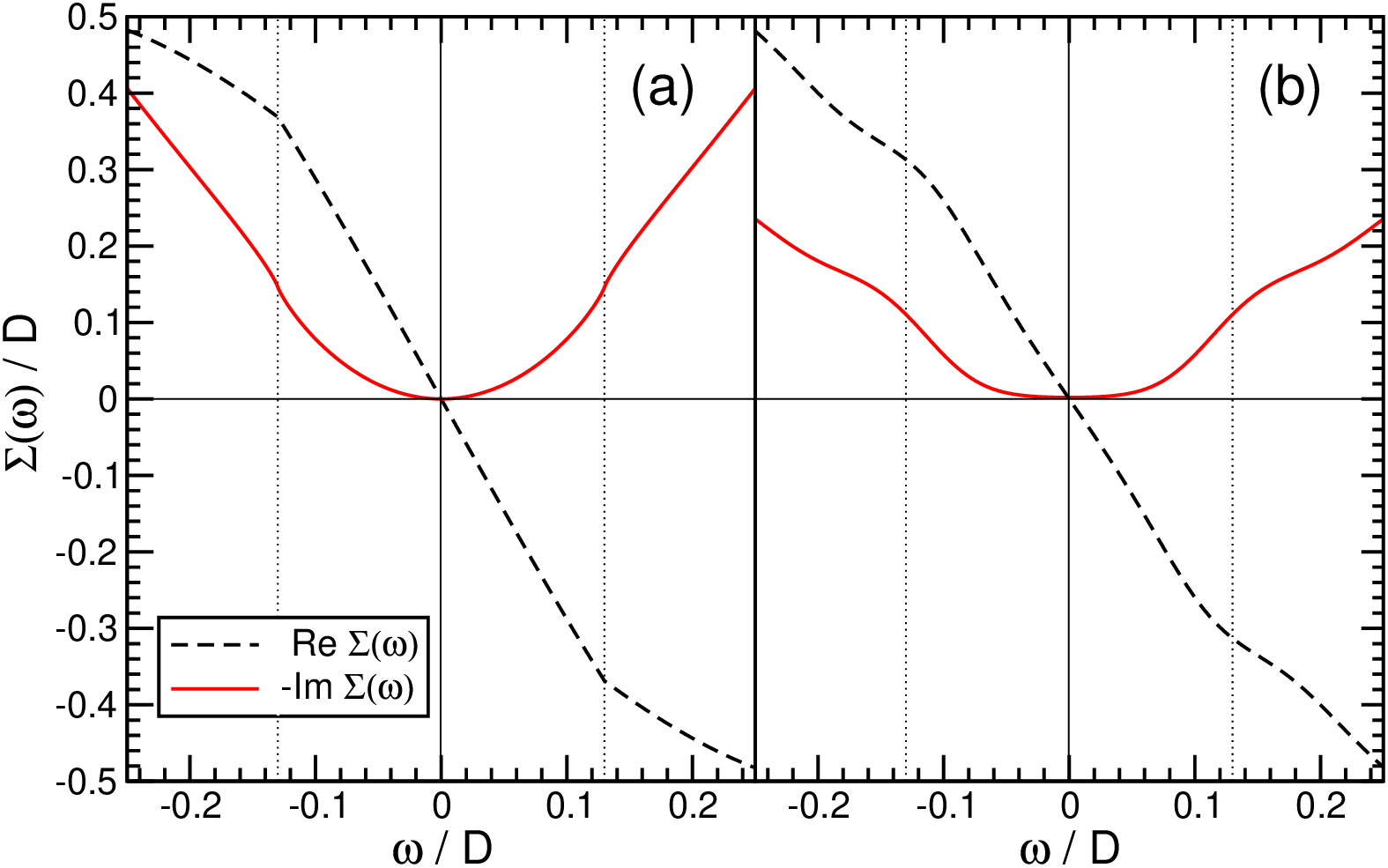}
\end{center}
\caption{(Color online) Panel (a) illustrates the relation between
real and imaginary part for an analytic ansatz of the real part with 
a kink: 
$\text{Re}\Sigma=\frac{-A^2}{2}\frac{(2\omega+|\omega-a|-|\omega+a|)
(\omega^2-b^2)}{A^2+\omega^4}$ with $A\approx 0.62177 D, a=0.15D, b=0.7D$;
the imaginary part is computed by the Kramers-Kronig relation. Panel (b) 
shows the real and imaginary part of $\Sigma(\omega)$ in DMFT
at $U=2.0D$.
 \label{fig:trough}}
\end{figure}
\begin{figure}
\begin{center}
     \includegraphics[width=0.82\columnwidth]{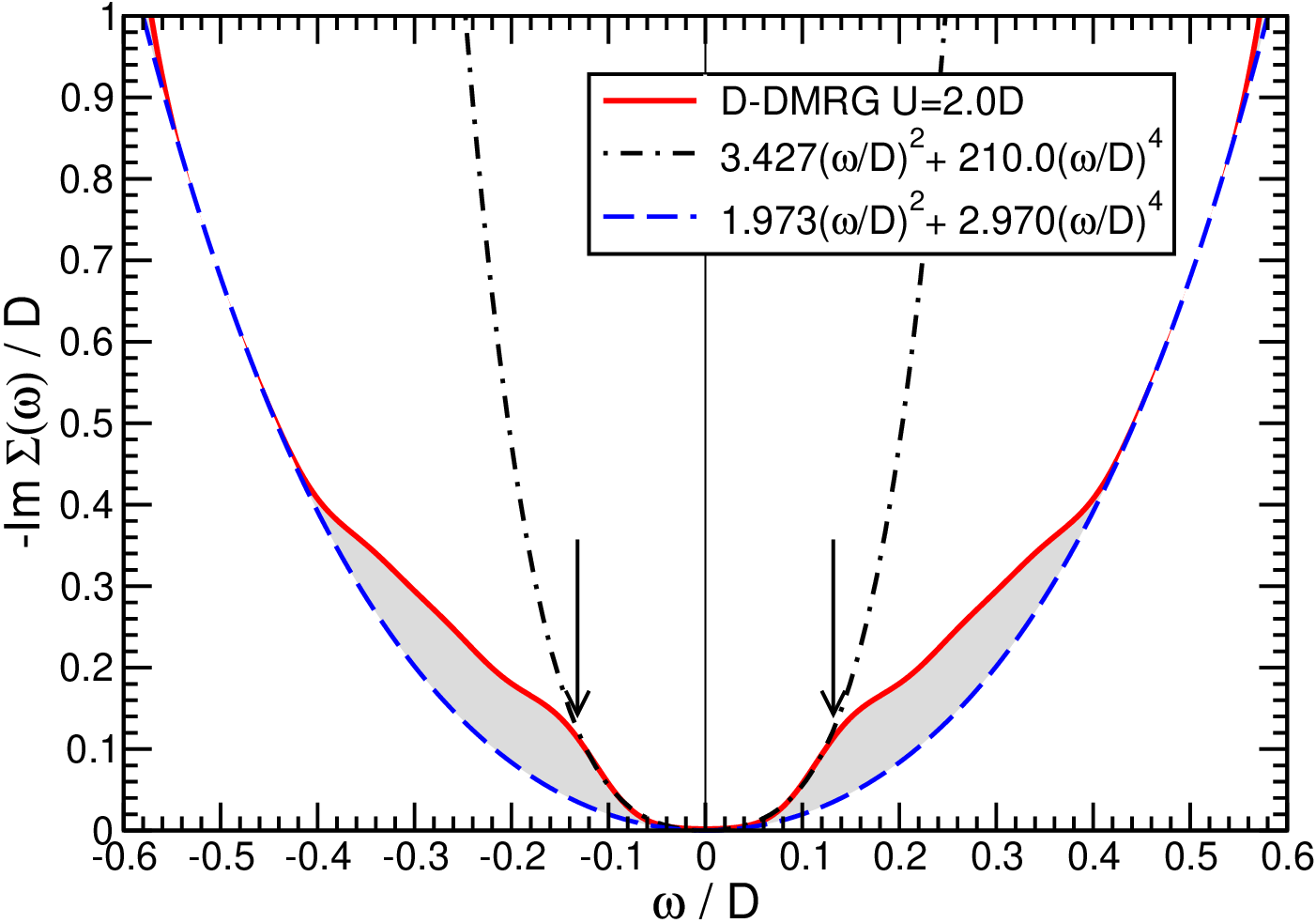}
\end{center}
\caption{(Color online)  $-\text{Im}\Sigma(\omega)$ on a larger
scale with two Fermi-liquid fits.
The shaded region illustrates  additional decay indicated by the arrows.
 \label{fig:selfenergy}}
\end{figure}

Figs.\ \ref{fig:trough} and \ref{fig:selfenergy} 
 display the generic behavior found
for $U$ not too far from the metal-insulator transition.
Fig.\ \ref{fig:trough}(a) shows that a kink in the 
real part of a self-energy is linked to a trough-like
feature in the imaginary part. This is a purely mathematical
fact stemming from Kramers-Kronig relation. 
Panel (b) depicts real and imaginary part for a realistic
self-energy as it results from the DMFT calculation. The kinks in the 
real part and the trough in the imaginary part are clearly discernible though 
not as neatly as in the analytic function of panel (a). This comes from
small spurious wiggles in $-\text{Im}\Sigma$ resulting inevitably from
the deconvolution of the DMRG raw data \cite{raas05a}. 

In Fig.\  \ref{fig:selfenergy} 
we address the physical meaning of the trough-like feature. 
There are two ways to understand it based on Fermi liquid
theory. 

(i) The trough itself, ranging 
approximately from $-0.1D$ to $0.1D$, 
is fitted by a narrow curve (dashed-dotted line). 
Outside the trough $-\text{Im}\Sigma(\omega)$ is then much lower than
the extrapolated fit. 
Since $-\text{Im}\Sigma(\omega)$ is the decay rate
for the quasiparticles,  this would be much lower relative to its  
extrapolated value.
We do not see a good reason for such a decrease of the decay because 
the decrease due to reduced phase space for three quasi-particles 
should occur beyond about three times $ZD/2$ which is $ \approx 0.4D$ 
\cite{bulla99a,karsk05,karsk08}, i.e., significantly larger than
the extension of the trough.

(ii) The Fermi liquid theory extends to higher values, for
instance $\approx 0.5D$ as is still consistent with the above crude estimate, 
so that fits such as the dashed one in Fig.\ \ref{fig:selfenergy} are 
justified. Indeed, the fit works very nicely with a moderate coefficient
for the quartic term. This view implies that around $0.15D$ \emph{additional}
decay becomes possible which extends up to $0.4D$. 
An additional decay channel is well possible. It sets in only 
above a certain energy because excitations of a certain minimum 
energy are involved.

So, among the two hypotheses we favor the second one. 
It explains also rather naturally why the quadratic coefficient is so low in 
spite of the very narrow trough. 

We are aware that from a puristic
point of view on Fermi liquid theory its applicability ends
at the borders of the narrow trough as described above in (i).
We do not claim that this view is invalid.
But we advocate the alternative view (ii) 
because it provides an intuitive way to understand the
self-energy behavior at low energies in terms of quasi-particles coupled to
emergent collective modes. This coupling is the origin for the
deviations from the dashed curve in Fig.\ \ref{fig:selfenergy}.
The latter is regarded as an effect on top of the underlying Fermi liquid
description resulting from the additional decay channel.
This is induced by scattering from an emergent collective mode
which has to be identified.

Given the fact that
it becomes important only for finite, though small energies we 
aim for a mode which is dominated by such a finite, though small energy.
Furthermore, it exists only close to the metal-insulator transition.
We shall see that its energy decreases towards the transition $U\to U_{c2}$.
Since the insulator is a paramagnet with disordered local spin moments
\cite{georg96,raas09} a natural candidate are the emergent
spin fluctuations. 

In the framework of the limit of infinite dimension
$d\to\infty$ the propagation of a collective mode from site $i$ to site
$j$ scales like $d^{-|i-j|}$ where $|\cdot|$ stands for the taxi cab
metric. Hence the collective modes are almost dispersionless and
thus local. Only for particular wave vectors which are of measure zero
a non-local propagation makes itself felt, see, e.g., Ref.\ 
\onlinecite{uhrig93b}. In the complex diagrams describing the
single particle motion the propagation of collective modes 
(particle-hole pairs) occurs in such a way that it is summed over.
No particular momenta of measure zero matter. Thus
 it is fully sufficient to investigate the local response.

\begin{figure}
\begin{center}
     \includegraphics[width=0.82\columnwidth]{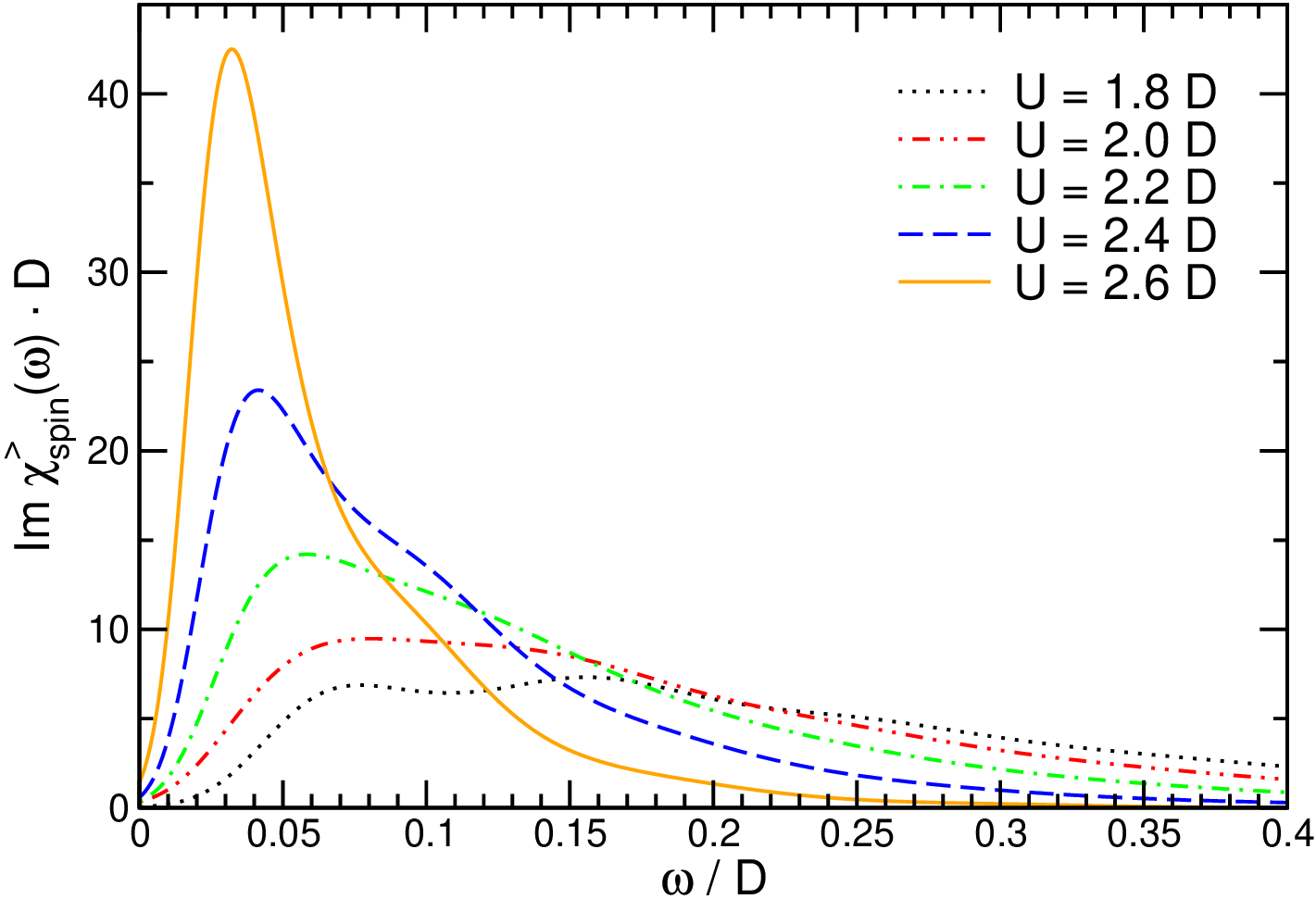}
\end{center}
\caption{(Color online) Deconvolved imaginary part of the local
spin susceptibility at positive frequencies for various interactions $U$ in
the metallic phase.}
 \label{fig:suscept}
\end{figure}
In Fig.\ \ref{fig:suscept} the local spin susceptibility 
$\chi_\text{spin}(\omega)$ is shown which we have computed
for positive frequencies  denoting it by $\chi^>_\text{spin}(\omega)$. It is
obtained from the effective SIAM
occurring in the self-consistency loop of the DMFT \cite{georg96,karsk08}.
In the SIAM it is determined as the local susceptibility at the head of the
 chain. Numerically we employ again D-DMRG for some broadening
$\eta$ which is then eliminated by deconvolution \cite{raas05a}.
This deconvolution gives rise to some uncertainty in the shape of the 
frequency dependence of the susceptibility.

The excitation operator is $2S^z=\hat n_\uparrow-\hat n_\downarrow$ 
at the chain head. A strongly pronounced peak catches the eye.
Its peak energy moves towards $\omega=0$ for $U\to U_{c2}$. In parallel, its
height increases such that its total weight tends to a finite value
\cite{raas09}. This peak is the precursor of a $\delta$ peak at zero
energy in the paramagnetic insulator. There it reflects the fact that 
a spin can be rotated without any cost of energy.
Still in the metallic phase, the peak is a resonance made from an almost
bound quasiparticle and a hole. It has some width
because it may decay into scattering states of its constituents.

\begin{figure}
\begin{center}
     \includegraphics[width=0.82\columnwidth]{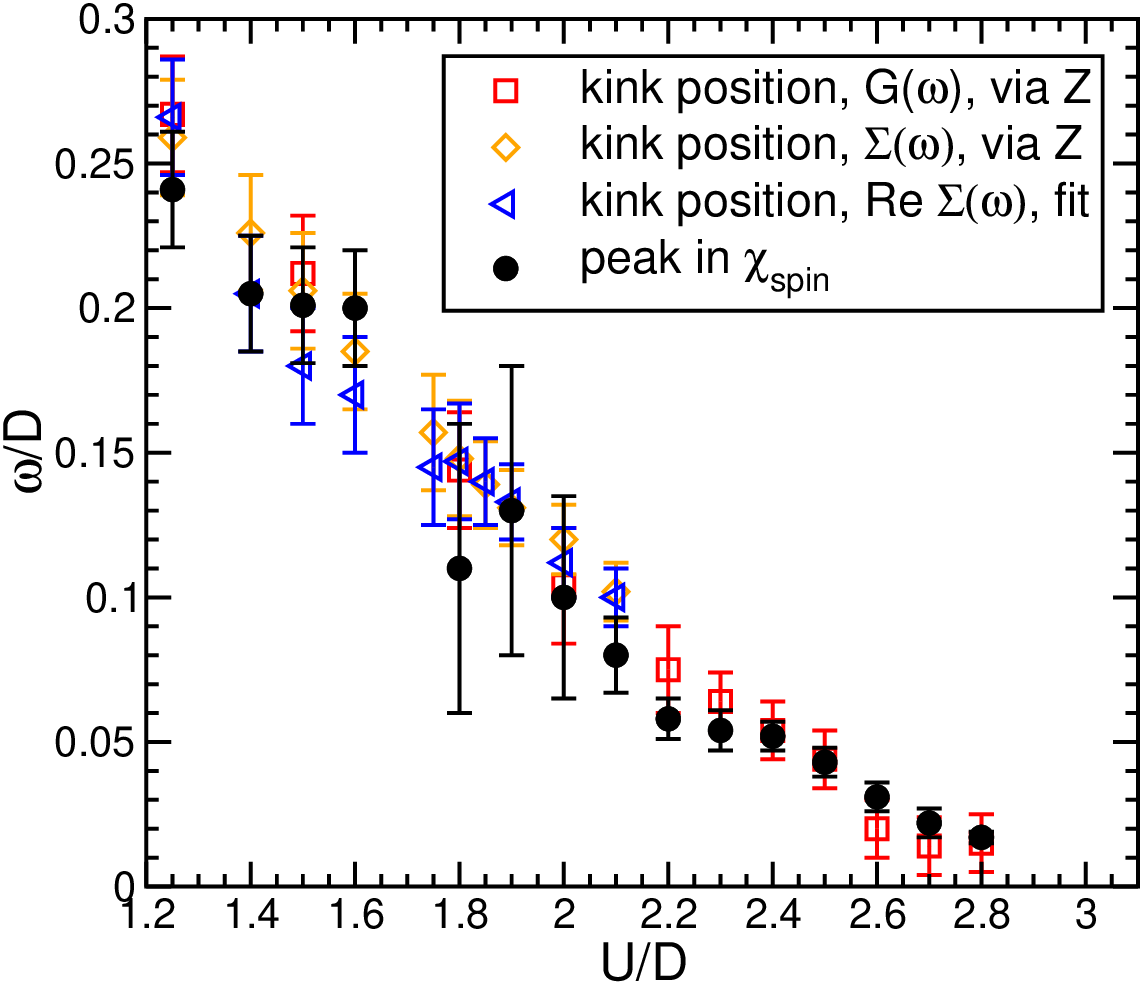}
\end{center}
\caption{(Color online) Kink positions $\omega_\text{kink}$
as derived from the quasi-particle 
weight $Z$ via Eq.\ (\ref{eq:byczuk}); $Z$ is found either from
the propagator $G$ \cite{karsk05} or from the self-energy $\Sigma(\omega)$
in $Z=(1-\partial_\omega \Sigma(0))^{-1}$. Most directly, a fit
$A \omega+B(|\omega-\omega_\text{kink}|-|\omega+\omega_\text{kink}|)$ to  
$\text{Re}\Sigma(\omega)$  is used
for $\omega_\text{kink}$; $\omega_\text{kink}$ is compared to
the energies where $\chi^>_{\text{spin}}$ 
shows a peak at low $|\omega|$.}
 \label{fig:energies}
\end{figure}
From the data for $\text{Im}\chi^>_{\text{spin}}$  
we deduce the peak position 
by fits assuming two lorentzians to account for the asymmetry of 
the peak shape. The lorentzians are 
multiplied by factors  $\tanh(\omega/\omega_0)$
to account for the linear vanishing of $\text{Im} \chi^>_\text{spin}(\omega)$ 
for $\omega\to 0$. The relevant peak position is the one of 
the lorentzian with more weight.
The error bars account for the uncertainties related
to the details of the fit procedure, e.g., for $U\approx 1.9D$
where the weight appears to be distributed equally over both lorentzians.

The results are compared in Fig.\ \ref{fig:energies} with the
kink positions which were determined in several ways. We use Eq.\
\eqref{eq:byczuk} to deduce the kink position from $Z$ which in turn
is determined either by $Z=(1-\partial_\omega \Sigma(0))^{-1}$
or by $Z^{-1}=D^2\partial_\omega G(0)/2$ \cite{karsk05}. Or the kink
position is determined directly by a fit to $\text{Re}\Sigma(0)$
(see caption of Fig.\ \ref{fig:energies}). The three ways to find the 
kink positions agree very well providing consistent data.

The peak positions agree remarkably well with the kink positions.
In particular for large values of $U$ the agreement is striking.
It is for these larger values $U\gtrapprox 2D$ that
both the kink and the peak in the susceptibility are clearly
discernible and well-defined.
So we deduce that the additional decay channel seen in Fig.\
\ref{fig:selfenergy} results from the excitation of
the spin resonance by the propagating single fermionic quasiparticle.
This finding strongly supports our claim that the kink is 
in fact due to emergent \emph{internal} modes. Here these modes are
the spin fluctuations which develop already in the metallic phase.

Thereby, an intuitive  physical picture of the origin of the
kinks is found.
One major advantage of this picture is that one can
transfer it to finite dimensions where the collective modes 
are dispersive so that the momentum dependence matters. Kinks are to be
expected where  momentum and energy conservation for the scattering
of a quasiparticle from a collective mode is fulfilled.

In conclusion, we have provided compelling evidence for a link between
the kinks in fermionic dispersions in strongly correlated systems
and emergent \emph{internal} collective modes, spin fluctuations
in particular. 
We agree completely with the phenomenon established by Byczuk et
al. \cite{byczu07a}. But our physical picture of the phenomenon
is different because we view the kinks as the consequence of 
inherent bosonic modes. An important concomitant aspect is that
the Fermi liquid theory does not break down already at the scale
of $\omega_\text{kink}$. It extends to about $2 ZD$ where
$Z$ is the quasiparticle weight. 

Our finding provides important information on the
possible interpretation of kinks in electronic dispersions
in many strongly correlated systems and in cuprate systems 
as they occur in high-temperature superconductors in particular.
Such kinks can  be the consequence of emerging bosonic modes,
i.e., resonances even if these are still strongly damped.
For instance, qualitative support is provided to
results based on the fluctuation-exchange approximation for cuprates 
\cite{mansk01}
Moreover, the coupling between the single particles and the collective modes
is generically substantial. Certainly, further investigations, for
instance away from half-filling, are called for.

\acknowledgments

We would like to thank M.\ Karski for providing data, H.\ Eschrig,
M.\ Kollar, I.\,A.\ Nekrasov, and D.\ Vollhardt for helpful discussions,
and the Heinrich Hertz-Stiftung NRW for financial support. 


\enlargethispage{3ex}

\end{document}